\documentclass[a4paper,twoside,11pt]{article} 

\usepackage[english]{babel} %
\usepackage{lmodern}
\usepackage[T1]{fontenc}
\usepackage[latin9]{inputenc}	

\newlength{\defaultparindent}
\usepackage{latexsym}
\usepackage{amsmath}
\usepackage{amsfonts}
\usepackage{amsthm}
\usepackage{amssymb}
\usepackage{bbold}
\usepackage{multirow}
\usepackage{hyperref}
\hypersetup{colorlinks=true,citecolor=green,linkcolor= blue}
\usepackage{todonotes} 
\usepackage{soul} 
\usepackage{tikz-cd}

\usepackage{algorithm} 
\usepackage{algpseudocode}

\usepackage[arXiv]{optional} 
\def\mynote{\todo} 

\opt{AACA}{
\def\cal{\mathcal}
}

\opt{x,AACA}{
\input{mbh_defs_TeX}
}

\opt{arXiv,std,JMP,JOPA}{

\newtheorem{MS_theorem}{Theorem}
\newtheorem{MS_lemma}{Lemma}
\newtheorem{MS_Proposition}{Proposition}

\def\myconjugate#1{\overline{#1}} 

\def\eg{e.g.\ }
\def\myisom{\cong} 
\newcommand{\R}{\ensuremath{\mathbb{R}}} 
\newcommand{\Identity}{\ensuremath{\mathbb{1}}} 

\def\my_span#1{\mbox{Span}\left(#1\right)} 

\def\dotinformula{\;\; \mathrm{.}} 
\def\OO#1{\ensuremath{\mbox{O}\!\left(#1\right)}}
\def\SO#1{\ensuremath{\mbox{SO}\!\left(#1\right)}}
\def\GL#1{\ensuremath{\mbox{GL}\!\left(#1\right)}}

\def\O1#1{\ensuremath{\mbox{O}^{#1}(1)}}
\newcommand{\comm}[2]{\ensuremath{\left[ #1, #2 \right]}}
\newcommand{\anticomm}[2]{\ensuremath{\left\{ #1, #2 \right\}}} 
\newcommand{\myClg}[3]{\ensuremath{{{\cal C}\ell} {\left( #3 \right)}}}	

}

\opt{JMP}{
}

\def\h_eigen{\eta}
\def\g_eigen{\theta}
\def\mygen{e} 
\def\myprimidemp{\mathbb{p}} 
\def\myprimidempset{\mathbb{P}} 

\def\SAT{\ensuremath{\mbox{SAT}}}
\def\bigO#1{\ensuremath{\mathcal{O}\left(#1\right)}}
\def\literal{\ensuremath{\rho}}
\def\mysetS{\mathcal{I}} 
\def\mySpinorS{{\mathbb{S}}} 
\def\myFockB{{\mathbb{F}}} 
\def\mysetM{\ensuremath{{{\cal M}_n}}} 
\def\mysnqG{\ensuremath{{{\cal N}_n}}} 
\def\BooT{\mathrm{T}} 
\def\BooF{\mathrm{F}} 
\def\Boov{Boolean variable} 
\def\BooS{{\cal{S}}} 

\begin{document}

\opt{x,std,arXiv,JMP,JOPA}{
\title{{\bf The Clifford algebra of $\R^{n,n}$ and the Boolean Satisfiability Problem} 
	}

\author{\\
	\bf{Marco Budinich}%
%
%
\\
	University of Trieste and INFN, Trieste, Italy\\ 
	\texttt{mbh@ts.infn.it}\\
%
%
%
	}
\date{ \today }
\maketitle
}

\vspace*{-5mm}
\begin{abstract}
We formulate a Boolean algebra in the set of idempotents of Clifford algebra $\myClg{}{}{\R^{n,n}}$ and within this frame we examine different formulations of the Boolean Satisfiability Problem in Clifford algebra. Exploiting the isomorphism between null subspaces of $\R^{n,n}$ associated to simple spinors and the orthogonal group $\OO{n}$ we ultimately give a continuous formulation of the Boolean Satisfiability Problem within this group that opens unexplored perspectives.
\end{abstract}


\opt{x,std,arXiv,JMP,JOPA}{

}

\opt{AACA}{
\maketitle
}

\section{Introduction}
\label{sec_Introduction}
Clifford algebra is a remarkably powerful tool initially developed to deal with automorphisms of quadratic spaces \cite{Porteous_1981} that was subsequently successfully applied to many other fields \eg to deal with combinatorial problems \cite{Schott_Staples_2010, Budinich_2017}.%
\opt{margin_notes}{\mynote{mbh.note: solita frasetta su Cartan ?? to more than a cantury ago... Un sentiero poco battuto è quello...}}%

On the other hand the Boolean Satisfiability Problem (\SAT{}) is the progenitor of many combinatorial problems and surprisingly fits neatly in the Clifford algebra of $\R^{n,n}$, $\myClg{}{}{\R^{n,n}}$. It is not the only case in which methods used in physics meet with \SAT{}: also statistical mechanics applied to \SAT{} have obtained considerable successes \cite{Braunstein_2005}.

Since Clifford algebra epitomizes geometric properties of its linear space \SAT{} formulation in \myClg{}{}{\R^{n,n}} grants it also a geometric meaning.

In this paper we introduce first a solid foundation of Boolean algebra within \myClg{}{}{\R^{n,n}} and in this frame we examine different formulations of \SAT{} from a unified standpoint. Being \SAT{} central to the $P \stackrel{\rm ?}{=} NP$ millennium problem \cite{Cook_2000} it is legitimate to ask whether this formulation can foster computational efficiency. Although only future will tell the neatness of this encoding adds a geometric meaning to Boolean expressions and conversely attach a Boolean significance to Clifford algebra idempotents.

In sections~\ref{prop_SAT_basics} and \ref{sec_neutral_R_nn} we succintly recall the Boolean Satisfiability problem and the Clifford algebra \myClg{}{}{\R^{n,n}}. Section~\ref{sec_Boole_in_Cl} provides a sturdy formulation of Boolean algebra with idempotents of \myClg{}{}{\R^{n,n}} tailored to our needs while in section~\ref{sec_SAT_in_Cl} we apply these results to give a neat encoding of \SAT{} in \myClg{}{}{\R^{n,n}} together with an unsatisfiability condition. To prepare the \SAT{} formulation in a more geometric setting in section~\ref{sec_O(n)_MTNS} we present in simple formalism the isomorphism between the set of all totally null subspaces of maximal dimension $n$ of $\R^{n,n}$ and the group $\OO{n}$ that ultimately in section~\ref{sec_SAT_in_O(n)} transforms a \SAT{} problem into the problem of forming a cover for the continuous group $\OO{n}$ that gives a continuous formulation of \SAT{} that seems to be new and that could also offer advantages to calculate solutions.

For the convenience of the reader we tried to make this paper as elementary and self-contained as possible.

\section{The Boolean Satisfiability problem}
\label{prop_SAT_basics}
The Boolean Satisfiability Problem \cite[Section~7.2.2.2]{Knuth_2015} seeks an assignment of $n$ \Boov{}s $\literal_i \in \{ \BooT, \BooF \}$ (true, false), that makes $\BooT$, \emph{satisfies}, a given Boolean formula $\BooS$ put in Conjunctive Normal Form (CNF) \eg
\begin{equation}
\label{formula_SAT_std}
\BooS \equiv (\literal_1 \lor \myconjugate{\literal}_2) \land (\literal_2 \lor \literal_3) \land ( \myconjugate{\literal}_1 \lor \myconjugate{\literal}_3) \land ( \myconjugate{\literal}_1 \lor \myconjugate{\literal}_2 \lor \literal_3) \land (\literal_1 \lor \literal_2 \lor \myconjugate{\literal}_3)
\end{equation}
as a logical AND ($\land$) of $m$ \emph{clauses} ${\cal C}_j$, the expressions in parenthesis, each clause being composed by the logical OR ($\lor$) of $k$ or less \Boov{}s possibly preceded by logical NOT ($\lnot\literal_i$, $\myconjugate{\literal}_i$ for short). In (\ref{formula_SAT_std}) $n = 3, m = 5$ and $k = 3$. To underline the difference with algebraic equality $=$ in what follows we use $\equiv$ to represent logical equivalence, namely that for all possible values taken by the \Boov{}s the two expressions are equal. A \emph{solution} is either an assignment of $\literal_i$ that gives $\BooS \equiv \BooT$ or a proof that such an assignment does not exist and $\BooS \equiv \BooF$.

\SAT{} was the first combinatorial problem proven to be NP-complete \cite{Cook_1971}; in particular while the case of $k = 3$, 3\SAT, can be solved only in a time that grows exponentially with $n$, $2\SAT$ and $1\SAT$ problems can be solved in polynomial time, that is \emph{fast}.%

Using the distributive properties of the logical operators $\lor, \land$ any given $k\SAT$ $\BooS$ expands in a logical OR of up to $k^m$ terms each term being a $1\SAT$ problem namely a logical AND of $m$ \Boov{}s. Since $\literal_i \land \myconjugate{\literal}_i \equiv \BooF$ the presence of a variable together with its logical complement is a necessary and sufficient condition for making a $1\SAT$ formula $\BooF$, namely \emph{unsatisfiable}, and thus these terms can be omitted and so $k^m$ is just an upper bound to the number of terms. Conversely a satisfiable $1\SAT$ formula has only one assignment of its \Boov{}s that makes it $\BooT$ and that can be read scanning the formula; thus in the sequel we will freely use $1\SAT$ formulas for assignments.

The final expanded expression can be further simplified and reordered exploiting the commutativity of the logical operators $\lor, \land$ and the properties $\literal_i \land \literal_i \equiv \literal_i \lor \literal_i \equiv \literal_i$. All ``surviving'' terms of this expansion, the Disjunctive Normal Form (DNF), are $1\SAT$ terms, each of them representing an assignment that satisfies the problem. On the contrary if the DNF is empty, as happens for (\ref{formula_SAT_std}), this is a proof that there are no assignments that make the formula $\BooT$: the problem is unsatisfiable.

Expansion to DNF is a dreadful algorithm for solving \SAT{}: first of all the method is an overkill since it produces all possible solutions whereas one would be enough; in second place this brute force approach gives a running time proportional to the number of expansion terms \bigO{(k^{\frac{m}{n}})^n} whereas the present best $\SAT$ solvers \cite{PaturiPudlakSaksZane_2005} run in \bigO{1.307^n}. Nevertheless DNF plays a central role in the formulation of \SAT{} in Clifford algebra.
%

\section{$\R^{n,n}$ and its Clifford algebra}
\label{sec_neutral_R_nn}
We review some properties of $\R^{n,n}$ and of its Clifford algebra $\myClg{}{}{\R^{n,n}}$ that are at the heart of the following results.

$\myClg{}{}{\R^{n,n}}$ is isomorphic to the algebra of real matrices $\R(2^n)$ \cite{Porteous_1981} and this algebra is more easily manipulated exploiting the properties of its Extended Fock Basis (EFB, see \cite{Budinich_2016} and references therein) with which any algebra element is a linear superposition of simple spinors. The $2 n$ generators of the algebra $\mygen_{i}$ form an orthonormal basis of the linear space $\R^{n,n}$
\begin{equation}
\label{formula_generators}
\mygen_i \mygen_j + \mygen_j \mygen_i := \anticomm{\mygen_i}{\mygen_j} = 2 \left\{ \begin{array}{l l}
\delta_{i j} & \mbox{for} \; i \le n \\
- \delta_{i j} & \mbox{for} \; i > n
\end{array} \right.
\qquad i,j = 1,2, \ldots, 2 n
\end{equation}
and we define the Witt, or null, basis of $\R^{n,n}$:
\begin{equation}
\label{formula_Witt_basis}
\left\{ \begin{array}{l l l}
p_{i} & = & \frac{1}{2} \left( \mygen_{i} + \mygen_{i + n} \right) \\
q_{i} & = & \frac{1}{2} \left( \mygen_{i} - \mygen_{i + n} \right)
\end{array} \right.
\quad i = 1,2, \ldots, n
\end{equation}
that, with $\mygen_{i} \mygen_{j} = - \mygen_{j} \mygen_{i}$ for $i \ne j$, gives
\begin{equation}
\label{formula_Witt_basis_properties}
\anticomm{p_{i}}{p_{j}} = \anticomm{q_{i}}{q_{j}} = 0
\qquad
\anticomm{p_{i}}{q_{j}} = \delta_{i j}
\end{equation}
showing that all $p_i, q_i$ are mutually orthogonal, also to themselves, that implies $p_i^2 = q_i^2 = 0$ and are thus null vectors.
\opt{margin_notes}{\mynote{mbh.note: should I mention here $\R^{2 n}_{hb}$ ? See book 780}}%
Defining
\begin{equation}
\label{formula_Witt_decomposition}
\left\{ \begin{array}{l}
P = \my_span{p_1, p_2, \ldots, p_n} \\
Q = \my_span{q_1, q_2, \ldots, q_n}
\end{array} \right.
\end{equation}
$P$ and $Q$ are two totally null subspaces of maximum dimension $n$ and form a Witt decomposition \cite{Porteous_1981} of $\R^{n,n}$ since $P \cap Q = \{ 0 \}$ and $P \oplus Q = \R^{n,n}$.

The $2^{2 n}$ simple spinors forming EFB are given by all possible sequences
\begin{equation}
\label{formula_EFB_def}
\psi = \psi_1 \psi_2 \cdots \psi_i \cdots \psi_n \qquad \psi_i \in \{ q_i p_i, p_i q_i, p_i, q_i \} \qquad i = 1, 2, \ldots, n
\end{equation}
where each $\psi_i$ takes one of its $4$ possible values \cite{Budinich_2016} and
\opt{margin_notes}{\mynote{mbh.note: should I use old version with $\mygen_{2 i - 1} \mygen_{2 i} = \comm{q_i}{p_i}$ ?}}%
each $\psi_i$ is uniquely identified by two ``bits'' $h_i, g_i = \pm1$: $h_i = 1$ if the leftmost vector of $\psi_i$ is $q_i$, $-1$ otherwise; $g_i = 1$ if $\psi_i$ is even, $-1$ if odd. The $h$ and $g$ \emph{signatures} of $\psi$ are respectively the vectors $(h_1, h_2, \ldots, h_n)$ and $(g_1, g_2, \ldots, g_n)$.

Since $\mygen_{i} \mygen_{i + n} = q_i p_i - p_i q_i := \comm{q_i}{p_i}$ in EFB the identity $\Identity$ and the volume element $\omega$ (scalar and pseudoscalar) assume similar expressions \cite{Budinich_2016}:
\begin{equation}
\label{formula_identity_omega}
\begin{array}{l l l}
\Identity & := & \anticomm{q_1}{p_1} \anticomm{q_2}{p_2} \cdots \anticomm{q_n}{p_n} \\
\omega & := & \mygen_1 \mygen_2 \cdots \mygen_{2 n} = (-1)^\frac{n(n-1)}{2} \comm{q_1}{p_1} \comm{q_2}{p_2} \cdots \comm{q_n}{p_n}
\end{array}
\end{equation}
and since $\myClg{}{}{\R^{n,n}}$ is a simple algebra, the algebra identity is also the sum of its $2^n$ primitive (indecomposable) idempotents $\myprimidemp_i$ we gather in set $\myprimidempset$
\begin{equation}
\label{formula_identity_def}
\Identity = \sum_{i = 1}^{2^n} \myprimidemp_i 
\qquad \myprimidemp_i \in \myprimidempset \dotinformula
\end{equation}
Comparing the two expressions of $\Identity$ we observe that the full expansion of the anticommutators of (\ref{formula_identity_omega}) contains $2^n$ terms each term being one of the primitive idempotents \emph{and} a simple spinor. $\myprimidempset$ is thus a proper subset of EFB (\ref{formula_EFB_def}) and its elements are
\begin{equation}
\label{formula_primitive_idempotents_def}
\myprimidemp = \psi_1 \psi_2 \cdots \psi_i \cdots \psi_n \qquad \psi_i \in \{ q_i p_i, p_i q_i \} \qquad i = 1, 2, \ldots, n \dotinformula
\end{equation}
We recall the standard properties of primitive idempotents
\begin{equation}
\label{formula_primitive_idempotents}
\myprimidemp_i^2 = \myprimidemp_i \quad (\Identity - \myprimidemp_i)^2 = \Identity - \myprimidemp_i \quad \myprimidemp_i (\Identity - \myprimidemp_i) = 0 \quad \myprimidemp_i \myprimidemp_j = \delta_{i j} \myprimidemp_i
\end{equation}
and define the set
\begin{equation}
\label{formula_S_def}
\mysetS := \left\{ \sum_{i = 1}^{2^n} \delta_i \myprimidemp_i : \delta_i \in \left\{0, 1 \right\}, \myprimidemp_i \in \myprimidempset \right\}
\end{equation}
in one to one correspondence with the power set of $\myprimidempset$. $\mysetS$ is closed under Clifford product but not under addition and is thus not even a subspace. With (\ref{formula_primitive_idempotents}) we easily prove
\begin{MS_Proposition}
\label{prop_S_properties}
For any $s \in \mysetS$ then $s^2 = s$.
\end{MS_Proposition}
\noindent $\mysetS$ is thus the set of the idempotents, in general not primitive; a simple consequence is that for any $s \in \mysetS$ also $(\Identity - s) \in \mysetS$.

\bigskip

Any EFB element of (\ref{formula_EFB_def}) is a simple spinor, uniquely identified in EFB by its $h$ signature, while the minimal left ideal, or spinor space $\mySpinorS$, to which it belongs is identified by its $h \circ g = (h_1 g_1, h_2 g_2, \ldots, h_n g_n)$ signature \cite{Budinich_2016}. The algebra, as a linear space, is the direct sum of these $2^n$ spinor spaces that, in isomorphic matrix algebra $\R(2^n)$, are usually associated to linear spaces of matrix columns.

For each of these $2^n$ spinor spaces $\mySpinorS$ its $2^n$ simple spinors (\ref{formula_EFB_def}) \cite{Budinich_2016} form a Fock basis $\myFockB$ and any spinor $\Psi \in \mySpinorS$ is a linear combination of the simple spinors $\psi \in \myFockB$ \cite{BudinichP_1989, Budinich_2016}. We illustrate this with the simplest example in $\R^{1,1}$, the familiar Minkowski plane of physics, here $\myClg{}{}{\R^{1,1}} \myisom \R(2)$ and the EFB (\ref{formula_EFB_def}) is formed by just 4 elements: $\{ qp_{+ +}, pq_{- -}, p_{- +}, q_{+ -} \}$ with the subscripts indicating respectively $h$ and $h \circ g$ signatures that give the binary form of the integer matrix indexes; its EFB matrix is
\begin{equation*}
\label{formula_EFB_1_1}
\bordermatrix{& + & - \cr
+ & q p & q \cr
- & p & p q \cr }
\end{equation*}
and, as anticipated, we can write the generic element $\mu \in \myClg{}{}{\R^{1,1}}$ in EFB
$$
\mu = \xi_{+ +} qp_{+ +} + \xi_{- -} pq_{- -} + \xi_{- +} p_{- +} + \xi_{+ -} q_{+ -} \qquad \xi \in \R \dotinformula
$$
The two columns are two minimal left ideals namely two (equivalent) spinor spaces $\mySpinorS_{+}$ and $\mySpinorS_{-}$. The two elements of each column are the simple spinors of Fock basis $\myFockB$ while $qp$ and $pq$ are the primitive idempotents and $qp + pq = \Identity$.

In turn simple spinors of a Fock basis $\myFockB$ are in one to one correspondence with null subspaces of maximal dimension $n$ of $\R^{n,n}$. For any $\psi \in \myFockB$ we define its associated maximal null subspace $M(\psi)$ as
\begin{equation}
\label{formula_MTNS_M_psi}
M(\psi) = \my_span{x_1, x_2, \ldots, x_n} \quad x_i = \left\{ \begin{array}{l l}
p_i & \mbox{iff} \; \psi_i = p_i, p_i q_i \\
q_i & \mbox{iff} \; \psi_i = q_i, q_i p_i
\end{array} \right.
\quad i = 1,2, \ldots, n
\end{equation}
\noindent and $x_i$ is determined by the $h$ signature of $\psi$ in EFB \cite{BudinichP_1989, Budinich_2016}. For example in $\myClg{}{}{\R^{3,3}}$ given the simple spinor $\psi = p_1 q_1 \, q_2 p_2 \, q_3 p_3$
$$
\psi = p_1 q_1 \, q_2 p_2 \, q_3 p_3 \quad \implies \quad M(\psi) = \my_span{p_1, q_2, q_3}
$$
and with (\ref{formula_Witt_basis_properties}) we see that for any $v \in M(\psi)$ then $v \psi = 0$.

We gather these $2^n$ maximal null subspaces of $\R^{n,n}$ in set \mysetM{} each of its elements being the span of the $n$ null vectors obtained choosing one null vector from each couple $(p_i, q_i)$ (\ref{formula_Witt_basis}). The set \mysetM{} is the same for all the $2^n$ different possible spinor spaces being identified by the $h$ signatures of $\psi$ in EFB and so it can be defined also starting from primitive idempotents (\ref{formula_primitive_idempotents_def}) and in summary we have three equivalent definitions for \mysetM{}
\begin{equation}
\label{formula_Mn}
\mysetM = \left\{ \begin{array}{l l}
\{ M(\psi) : \psi \in F \} \\
\{ \my_span{x_1, x_2, \ldots, x_n} : x_i \in \{ p_i, q_i \} \} \\
\{ M(\myprimidemp) : \myprimidemp \in \myprimidempset \} \dotinformula
\end{array} \right.
\end{equation}

\section{The Boolean algebra of \myClg{}{}{\R^{n,n}}}
\label{sec_Boole_in_Cl}
We exploit the known fact that in any associative, unital, algebra every family of commuting, orthogonal, idempotents generates a Boolean algebra to prove that the $2^{2^n}$ idempotents of $\mysetS$ (\ref{formula_S_def}) form a Boolean algebra.

A finite Boolean algebra is a set equipped with the inner operations of logical AND, OR and NOT that satisfy well known properties
but we will use an axiomatic definition \cite{Huntington_1933, Givant_Halmos_2009} that needs only a binary and a unary inner operations satisfying three simple axioms to prove:
\begin{MS_Proposition}
\label{prop_Boolean_algebra}
The set $\mysetS$ equipped with the two inner operations
\begin{equation}
\label{formula_S_properties}
\begin{array}{llll}
\mysetS \times \mysetS \to \mysetS & s_1, s_2 & \to & s_1 s_2 
\\
\mysetS \to \mysetS & s & \to & \Identity - s \\
\end{array}
\end{equation}
is a finite Boolean algebra.
\end{MS_Proposition}
\begin{proof}
We already observed that $\mysetS$ is closed under operations (\ref{formula_S_properties}) that moreover satisfy Boolean algebra axiomatic definition \cite{Huntington_1933}: the binary operation is associative since Clifford product is and commutative because all $\mysetS$ elements commute. The third (Huntington's) axiom requires that for any $s_1, s_2 \in \mysetS$
$$
(\Identity - (\Identity - s_1) s_2) (\Identity - (\Identity - s_1) (\Identity - s_2)) = s_1
$$
that is easily verified.
\end{proof}
\noindent We remark that $\mysetS$ is not a subalgebra of \myClg{}{}{\R^{n,n}} since it is not closed under addition. Any finite Boolean algebra is isomorphic to the power set of its Boolean atoms \cite{Givant_Halmos_2009}. In this case $\mysetS$ elements are in one to one correspondence with the power set of $\myprimidempset$ and we thus identify the Boolean atoms with the $2^n$ primitive idempotents (\ref{formula_primitive_idempotents_def}).

With simple manipulations we get all Boolean expressions in \myClg{}{}{\R^{n,n}}: in the unary operation of (\ref{formula_S_properties}) we recognize the logical NOT and associating the logical AND to Clifford product from $s (\Identity - s) = 0 \in \mysetS$ we deduce that $0$ stands for $\BooF$ and consequently that $\Identity$ stands for $\BooT$. For the logical OR we use De Morgan's relations
$$
\literal_i \lor \literal_j \equiv \myconjugate{\myconjugate{\literal}_i \land \myconjugate{\literal}_j} \quad \to \quad \Identity - (\Identity - s_1) (\Identity - s_2) = s_1 + s_2 - s_1 s_2
$$
and we can easily verify that $\mysetS$ is closed also under this binary operation.

We formulate Boolean expressions in \myClg{}{}{\R^{n,n}} associating \Boov{}s $\literal_i$ to idempotents and we gather associations in this table where $p_i$ and $q_i$ are vectors of the Witt basis (\ref{formula_Witt_basis})
\begin{equation}
\label{formula_Boolean_substitutions}
\begin{array}{lll}
\BooF & \to & 0 \\
\BooT & \to & \Identity \\
\literal_i & \to & q_i p_i \\
\myconjugate{\literal}_i & \to & \Identity - q_i p_i = p_i q_i \\
\literal_i \land \literal_j & \to & q_i p_i \; q_j p_j \\
\literal_i \lor \literal_j & \to & q_i p_i + q_j p_j - q_i p_i \; q_j p_j \dotinformula
\end{array}
\end{equation}
For example given some simple Boolean expressions with (\ref{formula_Witt_basis_properties}) we easily verify
$$
\begin{array}{lll}
\literal_i \land \literal_i \equiv \literal_i & \to & q_i p_i \; q_i p_i = q_i p_i \\
\myconjugate{\literal}_i \land \myconjugate{\literal}_i \equiv \myconjugate{\literal}_i & \to & p_i q_i \; p_i q_i = p_i q_i \\
\literal_i \land \myconjugate{\literal}_i \equiv \myconjugate{\literal}_i \land \literal_i \equiv \BooF & \to & q_i p_i \; p_i q_i = p_i q_i \; q_i p_i = 0 \\
\literal_i \land \literal_j \equiv \literal_j \land \literal_i & \to & q_i p_i \; q_j p_j = q_j p_j \; q_i p_i
\end{array}
$$
and from now on we will use $\literal_i$ and $\myconjugate{\literal}_i$ also in $\myClg{}{}{\R^{n,n}}$ meaning respectively $q_i p_i$ and $p_i q_i$ and Clifford product will stand for logical AND $\land$%
\opt{margin_notes}{\mynote{mbh.note: here there is a commented repetition of $1\SAT$ of section 2}}%
%
%
%
%
%
{} and in full generality we can prove \cite{Budinich_2017}
\begin{MS_Proposition}
\label{prop_logical_formulas_in_S}
Any Boolean expression $\BooS$ with $n$ \Boov{}s is represented in $\myClg{}{}{\R^{n,n}}$ by $S \in \mysetS$ obtained with substitutions (\ref{formula_Boolean_substitutions}) moreover $\myconjugate{\BooS}$ is represented by $\Identity - S$ both being idempotents of $\myClg{}{}{\R^{n,n}}$. Given another Boolean expression ${\cal Q}$ the logical equivalence $\BooS \equiv {\cal Q}$ holds if and only if $S = Q$ for their respective idempotents in $\myClg{}{}{\R^{n,n}}$.
\end{MS_Proposition}
\noindent In summary with substitutions (\ref{formula_Boolean_substitutions}) we can safely encode any Boolean expression, and thus \SAT{} problems, in Clifford algebra.

\section{\SAT{} in Clifford algebra \myClg{}{}{\R^{n,n}}}
\label{sec_SAT_in_Cl}

The straightest way of encoding a \SAT{} problem in CNF (\ref{formula_SAT_std}) in Clifford algebra is exploiting De Morgan relations to rewrite its clauses as
$$
{\cal C}_j \equiv (\literal_{j_1} \lor \literal_{j_2} \lor \cdots \lor \literal_{j_k}) \equiv \myconjugate{\myconjugate{\literal}_{j_1} \myconjugate{\literal}_{j_2} \cdots \myconjugate{\literal}}_{j_k}
$$
and thus the expression of a clause in Clifford algebra is
\begin{equation}
\label{formula_clause}
{\cal C}_j \to \Identity - \myconjugate{\literal}_{j_1} \myconjugate{\literal}_{j_2} \cdots \myconjugate{\literal}_{j_k} := \Identity - z_j
\end{equation}
and the expression of a \SAT{} problem in CNF with $m$ clauses is
\begin{equation}
\label{formula_SAT_EFB_2}
S = \prod_{j = 1}^m (\Identity - z_j)
\end{equation}
and from Proposition~\ref{prop_logical_formulas_in_S} easily descends
\begin{MS_Proposition}
\label{prop_SAT_in_Cl_2}
Given a \SAT{} problem $\BooS$ then $\BooS \equiv \BooF$ if and only if, for the corresponding algebraic expression in $\myClg{}{}{\R^{n,n}}$ (\ref{formula_SAT_EFB_2}) $S = 0$
\end{MS_Proposition}
\noindent that transforms a Boolean problem in an algebraic one. To master the implications of (\ref{formula_SAT_EFB_2}) we need the full expression of a $1\SAT$ formula \eg $\literal_1 \myconjugate{\literal}_2$ namely $q_1 p_1 \; p_2 q_2$: with (\ref{formula_identity_omega}), (\ref{formula_identity_def}) and (\ref{formula_primitive_idempotents_def})
\begin{equation}
\label{formula_literal_projection}
q_1 p_1 \; p_2 q_2 = q_1 p_1 \; p_2 q_2 \; \Identity = q_1 p_1 \; p_2 q_2 \prod_{j = 3}^{n} \anticomm{q_j}{p_j}
\end{equation}
since $q_1 p_1 \anticomm{q_1}{p_1} = q_1 p_1$ and $p_2 q_2 \anticomm{q_2}{p_2} = p_2 q_2$ and the full expansion of this expression is the sum of $2^{n - 2}$ primitive idempotents $\myprimidemp$ (\ref{formula_primitive_idempotents_def}) and thus $q_1 p_1 \; p_2 q_2$ is an idempotent of $\mysetS$. From the Boolean standpoint this can be interpreted as the property that given the $1\SAT$ formula $\literal_1 \myconjugate{\literal}_2$ the other, unspecified, $n-2$ \Boov{}s $\literal_3, \ldots, \literal_n$ can take all possible $2^{n - 2}$ values or, more technically, that $\literal_1 \myconjugate{\literal}_2$ has a \emph{full} DNF made of $2^{n - 2}$ Boolean atoms.

More in general any $1\SAT$ formula with $m$ \Boov{}s is a sum of $2^{n - m}$ primitive idempotents, namely Boolean atoms. With (\ref{formula_Boolean_substitutions}) we can rewrite (\ref{formula_primitive_idempotents_def}) as
$$
\myprimidemp = \psi_1 \psi_2 \cdots \psi_i \cdots \psi_n \qquad \psi_i \in \{ \literal_i, \myconjugate{\literal}_i \} \qquad i = 1, 2, \ldots, n
$$
showing that the $2^n$ primitive idempotents $\myprimidemp$ are just the possible $2^n$ $1\SAT$ formulas with $n$ \Boov{}s, the Boolean atoms, for example:
\begin{equation}
\label{formula_atoms_primitive}
\literal_1 \myconjugate{\literal}_2 \literal_3 \cdots \literal_n \to q_1 p_1 \; p_2 q_2 \; q_3 p_3 \cdots q_n p_n \in \myprimidempset \dotinformula
\end{equation}
%
%
By Proposition~\ref{prop_logical_formulas_in_S} $S \in \mysetS$ (\ref{formula_S_def}) and is thus the sum of primitive idempotents (\ref{formula_primitive_idempotents_def}) that now we know represent Boolean atoms and ultimately (\ref{formula_SAT_EFB_2}) gives the full DNF expansion of the \SAT{} problem $S$ each term being one assignment that makes the problem $\BooT$ while if the expansion is empty the problem is unsatisfiable and thus expansion of the CNF $S$ of (\ref{formula_SAT_EFB_2}) reproduces faithfully the Boolean expansion to DNF outlined in section~\ref{prop_SAT_basics}.

From the computational side Proposition~\ref{prop_SAT_in_Cl_2} is not a big deal since the expansion of (\ref{formula_SAT_EFB_2}) corresponds to the DNF expansion that in section~\ref{prop_SAT_basics} we named a ``dreadful'' algorithm. But porting \SAT{} to Clifford algebra offers other advantages since we can exploit algebra properties. For example the unsatisfiability condition $S = 0$ makes $S$ a scalar whereas if satisfiable $S$ is not a scalar. Exploiting scalar properties in Clifford algebra we proved \cite{Budinich_2017}
\begin{MS_theorem}
\label{theorem_SAT_unsat_thm}
A given nonempty \SAT{} problem in $\myClg{}{}{\R^{n,n}}$ (\ref{formula_SAT_EFB_2}) is unsatisfiable ($S = 0$) if and only if, for all generators (\ref{formula_generators}) of $\myClg{}{}{\R^{n,n}}$
\begin{equation}
\label{formula_SAT_symmetry}
\mygen_i \; S \; \mygen_i^{-1} = S \qquad \forall \; 1 \le i \le 2 n \dotinformula
\end{equation}
\end{MS_theorem}

\noindent This result gives an unsatisfiability test based on the symmetry properties of its CNF expression $S$ (\ref{formula_SAT_EFB_2}). We remark that as far as computational performances are concerned an efficient unsatisfiability test would bring along also an efficient solution algorithm. Suppose the test (\ref{formula_SAT_symmetry}) fails and thus that $S$ is satisfiable, to get an actual solution we choose a \Boov, \eg $\literal_i$, and replace it with $\BooT$ and apply again the test to the derived problem $S_i$. If the test on $S_i$ fails as well this means that $\literal_i \equiv \BooT$ otherwise, necessarily, $\literal_i \equiv \BooF$ and repeating this procedure $n$ times for all \Boov{}s we obtain an assignment that satisfies $S$. The algorithmic properties of unsatisfiability test (\ref{formula_SAT_symmetry}) have been preliminarly explored in \cite{Budinich_2017}.

\bigskip

$\myClg{}{}{\R^{n,n}}$ epitomizes the geometry of linear space $\R^{n,n}$ and thus \SAT{} encoding (\ref{formula_SAT_EFB_2}) brings along also a geometric interpretation that is at the root of further encodings of \SAT{} in Clifford algebra.

$S$ (\ref{formula_SAT_EFB_2}) is ultimately a sum of primitive idempotents (\ref{formula_primitive_idempotents_def}) that are in one to one correspondence with the null maximal subspaces of \mysetM{} (\ref{formula_Mn}).

It follows that $S$, and more in general any $\mysetS$ element, induces a subset of \mysetM{}, the empty subset if $S = 0$. More precisely the elements of this subset are all and only those maximal totally null subspaces (\ref{formula_Mn}) corresponding to the Boolean atoms making $S$. For any $s \in \mysetS$ (\ref{formula_S_def}) let $I_s$ such that
\begin{equation}
\label{formula_J_s_def}
s = \sum_{i \in I_s} \myprimidemp_i \qquad I_s \subseteq \{ 1, 2, \ldots 2^n \}
\end{equation}
and so all $s \in \mysetS$ induce a subset of \mysetM{}
\begin{equation}
\label{formula_cal_T_j_def}
{\cal T'}_s := \{M(\myprimidemp_i) : i \in I_s \} \subseteq \mysetM
\end{equation}
and we see immediately that
\begin{equation}
\label{formula_cal_T_j_complementary}
{\cal T'}_{\Identity - s} = \mysetM - {\cal T'}_s \dotinformula
\end{equation}
With these definitions Proposition~\ref{prop_SAT_in_Cl_2} takes a different form:
\begin{MS_Proposition}
\label{prop_SAT_in_M_n}
Given a \SAT{} problem $S$ in $\myClg{}{}{\R^{n,n}}$ (\ref{formula_SAT_EFB_2}) then the problem is unsatisfiable ($S = 0$) if and only if
\begin{equation}
\label{formula_SAT_in_M_n}
\cup_{j = 1}^m {\cal T'}_{z_j} = \mysetM \dotinformula
\end{equation}
\end{MS_Proposition}
\begin{proof}
For any $s_1, s_2 \in \mysetS$ from $\mysetS$ definition (\ref{formula_S_def}) we easily get
$$
{\cal T'}_{s_1 s_2} = {\cal T'}_{s_1} \cap {\cal T'}_{s_2}
$$
and in this setting Proposition~\ref{prop_SAT_in_Cl_2} $S$ states that $S = 0$ if and only if
\begin{equation}
\label{formula_SAT_in_T}
{\cal T'}_S = \cap_{j = 1}^m {\cal T'}_{\Identity - z_j} = \emptyset \dotinformula
\end{equation}
The thesis follows by (\ref{formula_cal_T_j_complementary}) and by elementary set properties.
\end{proof}
\noindent The \SAT{} problem has now the form of a problem of subsets of \mysetM{} that provides also an interpretation of (\ref{formula_SAT_EFB_2}). Since $z_j$ is the unique assignment of the $k$ \Boov{}s of ${\cal C}_j$ that give ${\cal C}_j \equiv \BooF$, and thus $S = 0$, then if the union of all these cases (\ref{formula_SAT_in_M_n}) covers \mysetM{} the problem is unsatisfiable. This in turn implies that any $M(\myprimidemp) \notin \cup_{j = 1}^m {\cal T'}_{z_j}$ is a solution of $S$.

To proceed further we review the isomorphism between the set of all totally null subspaces of maximal dimension of $\R^{n,n}$ and the group $\OO{n}$.

\section{The orthogonal group $\OO{n}$ and the set \mysnqG{}}
\label{sec_O(n)_MTNS}
Let \mysnqG{} be the set of all totally null subspaces of maximal dimension $n$ of $\R^{n,n}$, a quadric Grassmannian for Ian Porteous \cite[Chapter~12]{Porteous_1981}. \mysnqG{} is isomorphic to subgroup $\OO{n}$ of $\OO{n,n}$ and $\OO{n}$ acts transitively on \mysnqG{}.

We review these relations: seeing the linear space $\R^{n,n}$ as $\R^n \times \R^n$ we can write its generic element as $(x,y)$ and $(x,y)^2 = x^2 - y^2$. Any $n$ dimensional subspace of $\R^n \times \R^n$ may be represented as the image of an injective map $\R^n \to \R^n \times \R^n; x \to (u(x), s(x))$ for $u, s \in \GL{n}$. This subspace is made by all pairs $(u(x), s(x))$ and we denote it with $(u,s) \in \GL{n} \times \GL{n}$.

By same mechanism for any $x \in \R^n \times \{0\}$ and $t \in \OO{n}$ $(x,t(x))$ is a null vector of $\R^{n,n}$ since $(x,t(x))^2 = x^2 - t(x)^2 = 0$ and it belongs to the $n$ dimensional null subspace $(\Identity, t)$. Isometries $t \in \OO{n}$ establish the quoted isomorphism since any subspace $(\Identity, t) \subset \R^{n,n}$ is in \mysnqG{} and conversely any element of \mysnqG{} can be written as $(\Identity, t)$ \cite[Corollary~12.15]{Porteous_1981} and thus
\begin{equation}
\label{formula_Nn}
\mysnqG = \{ (\Identity, t) : t \in \OO{n} \}
\end{equation}
and the isomorphism between \mysnqG{} and $\OO{n}$ is realized by map
\opt{margin_notes}{\mynote{mbh.note: is this the Cayley chart ?}}%
\begin{equation}
\label{formula_bijection_Nn_O(n)}
\mysnqG \to \OO{n}; (\Identity, t) \to t \dotinformula
\end{equation}

For example assuming that the map $(\Identity, \Identity): \R^n \times \{0\} \to \R^n \times \R^n$ is such that $\mygen_i \to (\mygen_i, \mygen_{i + n})$ than two generic null vectors of $P$ and $Q$ (\ref{formula_Witt_decomposition}) are respectively $(x, x)$ and $(y, -y)$ and in this notation $P$ and $Q$ are thus
\opt{margin_notes}{\mynote{mbh.ref: see pp. 55, 56}}%
\begin{equation}
\label{formula_P_Q_def}
\left\{ \begin{array}{l}
P = (\Identity, \Identity) \\
Q = (\Identity, -\Identity) \dotinformula
\end{array} \right.
\end{equation}
The action of $\OO{n}$ is transitive on \mysnqG{} since for any $t, u \in \OO{n}$, $(\Identity,ut) \in \mysnqG$ and the action of $\OO{n}$ is trivially transitive on $\OO{n}$.

We examine isomorphism (\ref{formula_bijection_Nn_O(n)}) when restricted to subset $\mysetM \subset \mysnqG$ (\ref{formula_Mn}) and we take $P = (\Identity, \Identity)$ as our ``reference'' element of \mysetM{}.
\opt{margin_notes}{\mynote{mbh.note: in physics language this corresponds to the choice of the vacuum spinor}}%

Let $\lambda_i$ be the hyperplane reflection inverting timelike vector $\mygen_{i + n}$, namely
\begin{equation*}
\label{formula_t_i_def}
\lambda_i(\mygen_{j}) =
\left\{ \begin{array}{l l l}
-\mygen_{j} & \quad \mbox{for} \quad j = i + n\\
\mygen_{j} & \quad \mbox{otherwise}
\end{array} \right.
\quad i = 1,2, \ldots, n \quad j = 1,2, \ldots, 2n
\end{equation*}
its action on the Witt basis (\ref{formula_Witt_basis}) exchanges the null vectors $p_i$ and $q_i$. Starting from $P$ (\ref{formula_Witt_decomposition}) we can get any other \mysetM{} element inverting a subset of the $n$ timelike vectors $\mygen_{i + n}$. Each isometry $\lambda_i$ acts on the (timelike) subspace $\{0\} \times \R^n$ of $\R^{n,n}$ and is represented, in the vectorial representation of $\OO{n}$, by a diagonal matrix $\lambda \in \R(n)$ with $\pm 1$ on the diagonal and these matrices form the group
$$
\OO{1} \times \OO{1} \cdots \times \OO{1} = \stackrel{n}{\times} \OO{1} := \O1{n}
$$
immediate to get since $\OO{1} = \{ \pm 1 \}$.%
\opt{margin_notes}{\mynote{mbh.ref: remember that the tensor product is associative Proposition~11.5 of {Porteous 1995}; $\O1{n}$ is not normal since in general $t \lambda t^{-1} \notin \O1{n}$}}%
{} \O1{n} is a discrete, abelian, subgroup of involutions of $\OO{n}$, namely linear maps $t$ such that $t^2 = \Identity$. It is thus clear that since $P = (\Identity, \Identity)$ for any $M(\myprimidemp) \in \mysetM$ there exists a unique $\lambda \in \O1{n}$ such that
$$
M(\myprimidemp) = (\Identity, \lambda)
$$
and thus we proved constructively
\begin{MS_Proposition}
\label{prop_bijection_restricted}
Isomorphism (\ref{formula_bijection_Nn_O(n)}) restricted to $\mysetM \subset \mysnqG$ has for image subgroup \O1{n} of $\OO{n}$
\begin{equation}
\label{formula_bijection_Mn_O^n(1)}
\mysetM = \{ (\Identity, \lambda) : \lambda \in \O1{n} \} \qquad \implies \qquad \mysetM \to \O1{n}; (\Identity, \lambda) \to \lambda \dotinformula
\end{equation}
\end{MS_Proposition}

\noindent Given reference $P$ and simple spinor $\psi_P = \prod_{i = 1}^{n} p_i q_i$, such that $M(\psi_P) = P$, the vacuum spinor of physics, we resume concisely the action of $\lambda \in \O1{n}$ on spinors and vectors with (see \eg \cite{BudinichP_1989, Budinich_2016} for more extensive treatments)
\begin{equation}
\label{formula_lambda_action}
M(\lambda(\psi_P)) := M(\psi_\lambda) = (\Identity, \lambda)
\end{equation}
and by the action of $\lambda$ we get respectively from $\psi_P$ all spinors of the Fock basis $\myFockB$ and from null subspace $P$ all \mysetM{} elements. We resume all this in a commutative diagram in which numbers refer to formulas
$$
\begin{tikzcd}[column sep=small]
& \myprimidemp \in \myprimidempset \arrow[dl, Leftrightarrow, "(\ref{formula_Mn})" '] \arrow[dr, Leftrightarrow, "(\ref{formula_lambda_action})"] & \\
M(\myprimidemp)
\in \mysetM \arrow[Leftrightarrow, "(\ref{formula_bijection_Mn_O^n(1)})"]{rr} & & \lambda \in \O1{n} \dotinformula
\end{tikzcd}
$$

\section{\SAT{} in orthogonal group $\OO{n}$}
\label{sec_SAT_in_O(n)}
Isomorphism (\ref{formula_bijection_Mn_O^n(1)}) adds a fourth definition of \mysetM{} (\ref{formula_Mn}) with which we can port \SAT{} in group $\OO{n}$. We redefine ${\cal T'}_s$ (\ref{formula_cal_T_j_def}) as a subset of \O1{n}
\begin{equation}
\label{formula_cal_T_j_def2}
{\cal T'}_s := \{ \lambda \in \O1n : (\Identity, \lambda) = M(\myprimidemp_i), i \in I_s \}
\end{equation}
and with this definition we can transform Proposition~\ref{prop_SAT_in_M_n} to
\begin{MS_Proposition}
\label{prop_SAT_in_On1}
Given a \SAT{} problem $S$ in $\myClg{}{}{\R^{n,n}}$ (\ref{formula_SAT_EFB_2}) then the problem is unsatisfiable ($S = 0$) if and only if
\begin{equation}
\label{formula_SAT_in_On1}
\cup_{j = 1}^m {\cal T'}_{z_j} = \O1{n}
\end{equation}
\end{MS_Proposition}
\noindent that gives the first formulation of \SAT{} problems in group language. From the computational point of view of there are no improvements since $\O1{n}$ is a discrete group and checking if subsets ${\cal T'}_{z_j}$ form a cover essentially requires testing all $2^n$ group elements, just the same as testing all $2^n$ Boolean atoms to see if any solves \SAT{}.

In the last step we show that, when problem $S$ is unsatisfiable, subsets induced by clauses not only form a cover of \O1{n} (\ref{formula_SAT_in_On1}) but also of its parent group $\OO{n}$ that opens interesting computational perspectives.

We start showing that the same null subspace of \mysnqG{} may assume different forms $(\Identity, t)$, for example by a change of basis \cite{Budinich_2020}. We introduce the arguments
\opt{margin_notes}{\mynote{mbh.note: see log pp. 716 ff}}%
with two lemmas proving basic properties of \mysnqG{}:
\begin{MS_lemma}
\label{lemma_MTNS_equivalence}
Given any $u, s \in \GL{n}$ such that $s u^{-1} \in \OO{n}$ then $(u, s)$ and $(\Identity, s u^{-1})$ represent the same null subspace.
%
\opt{margin_notes}{\mynote{mbh.note: very similar to \cite[Proposition~12.14]{Porteous_1981}, for a fully general version see log. p. 721'}}%
\end{MS_lemma}
\begin{proof}
Since $u, s \in \GL{n}$ any $(u(a), s(a)) := (x, y)$ may be written as $(x, s u^{-1}(x))$, namely an element of $(\Identity, s u^{-1}) \in \mysnqG{}$ (\ref{formula_Nn}).
\end{proof}

In vectorial representations matrix $\Identity_n$ may be thought as the standard basis of $\R^n \times \{0\}$ and the column vectors of matrix $s u^{-1} \in \OO{n}$ are the action of $s u^{-1}$ on the standard basis since $s u^{-1} \Identity_n = s u^{-1}$. Subspace $(\Identity, s u^{-1})$ is thus the space spanned by $n$ column vectors $\left(\begin{array}{c} \Identity \\ s u^{-1} \end{array}\right)$ and we will switch freely between the two representations. In this view column vectors $\left(\begin{array}{c} u \\ s \end{array}\right)$ are just a linear combination of $\left(\begin{array}{c} \Identity \\ s u^{-1} \end{array}\right)$ by $u \in \GL{n}$ and this explains Lemma~\ref{lemma_MTNS_equivalence} and allows us to define $(u, s)$ and $(\Identity, s u^{-1})$ as two equivalent \emph{forms} representing the same subspace and we denote this with $(u, s) \simeq (\Identity, s u^{-1})$.

\begin{MS_lemma}
\label{lemma_MTNS_diagonalization}
Given $(\Identity, t_1), (\Identity, t_2) \in \mysnqG$ represented with $2 n$ column vectors $W = \left(\begin{array}{c c} \Identity & \Identity \\ t_1 & t_2 \end{array}\right)$ it is always possible to find an orthogonal basis of $\R^{n, n}$ in which $W$ becomes $W' = \left(\begin{array}{c c} \Identity & \Identity \\ \Identity & u^{T} t u \end{array}\right)$ with $t = t_1^T t_2$ and for any $u \in \OO{n}$.
\end{MS_lemma}
\begin{proof}
In an orthogonal basis of $\R^{n,n}$ with column vectors $B = \left(\begin{array}{c c} u & 0 \\ 0 & t_1 u \end{array}\right)$, with $u \in \OO{n}$, W becomes $W' = B^{-1} W = \left(\begin{array}{c c} u^{T} & u^{T} \\ u^{T} & u^{T} t_1^T t_2 \end{array}\right)$ and the thesis follows with Lemma~\ref{lemma_MTNS_equivalence}.
\end{proof}

\noindent This proves two facts: the first is that any $(\Identity, t_1) \in \mysnqG$ may be chosen as the reference element $P = (\Identity, \Identity)$ corresponding to the freedom of choosing an arbitrary vacuum spinor.

The second is that together with $P$ any other $(\Identity, t) \in \mysnqG$ can take all the forms $(\Identity, u^{T} t u)$ for any $u \in \OO{n}$, in particular it could be diagonal; we can extend our notation to $(\Identity, t) \simeq (\Identity, u^{T} t u)$ while $P$ remains $(\Identity, \Identity)$ in all orthogonal bases.
%
%
This confirms the intuition that whatever form takes $t \in \OO{n}$ in an orthogonal basis $(\Identity, t)$ always represents the same subspace.
Orthogonal similarity is an equivalence relation in $\OO{n}$ and each equivalence class represents one maximal totally null subspace. The elements of one class are different forms the same subspace assumes in different orthogonal bases.

We continue proving that \mysnqG{} contains only $2^n$ \emph{different} subspaces.
\begin{MS_Proposition}
\label{prop_2^n_MTNS}
The set \mysnqG{} of $\R^{n,n}$ contains only $2^n$ different subspaces corresponding, via bijection (\ref{formula_bijection_Mn_O^n(1)}), to the elements of $\O1{n}$; all other $t \in \OO{n}$ being just another form of one of these subspaces.
\end{MS_Proposition}
\begin{proof}
We proceed by induction on $n$, for $n = 1$ the proposition is true since $\OO{1} = \{ \pm 1 \}$ and bijection (\ref{formula_bijection_Nn_O(n)}) proves the thesis%
%
%
. Let the proposition be true for $n - 1$, moving to $n$ for any null vector $(x,y) \in \R^{n,n}$ then also $(x,-y)$ is a null vector and seeing $\R^{n,n}$ as $\R^n \times \R^n$ we can split respectively $\R^n \times \{0\} = \{x\} \oplus \{x\}^\perp$ and $\{0\} \times \R^n = \{y\} \oplus \{y\}^\perp$ and clearly $\{x\}^\perp \oplus \{y\}^\perp \myisom \R^{n-1,n-1}$ that by hypothesis contains $2^{n-1}$ null subspaces and any of this subspaces can be completed with either of the two null subspaces $(x,y), (x,-y) \in \{x\} \oplus \{y\} = \R^{1,1}$ that proves the thesis.
\opt{margin_notes}{\mynote{mbh.note: other possible proofs: by decomposition of Clifford algebra $\myClg{}{}{\R^{n,n}}$ \cite[p.~247]{Porteous_1981} or by decomposition of hyperbolic spaces: e.g. § 6 of paper 768 or § 42D of O'Meara book 447 (where else did I see this ?)}}%
\end{proof}
This property of \mysnqG{} appeared, albeit in different form, in the book of Élie Cartan \cite{Cartan_1937} but, after that, rarely surfaced again in the literature an exception being \cite{BudinichP_1989}. It establishes another equivalence relation on $\OO{n}$ \cite{Budinich_2020}

\begin{MS_Proposition}
\label{prop_equivalence_relation}
The binary relation on $\OO{n}$ $u \sim t$ if and only if both $(\Identity, u)$ and $(\Identity, t)$ represent the same null subspace $(\Identity, \lambda)$, with $\lambda \in \O1{n}$, defines an equivalence relation and its $2^n$ equivalence classes are
\begin{equation}
\label{formula_C_t_general}
C_\lambda := \left\{ t \in \OO{n} : (\Identity, t) \simeq (\Identity, \lambda), \lambda \in \O1{n}
 \right\} \dotinformula
\end{equation}
\end{MS_Proposition}

We can thus associate to a \mysetM{} element (\ref{formula_bijection_Mn_O^n(1)}) not only $\lambda \in \O1n$ but its whole equivalence class $C_\lambda$ representing it in all possible forms. Consequently we pass from ${\cal T'}_{z_j}$ of (\ref{formula_cal_T_j_def2}) to superset ${\cal T}_{z_j} \supset {\cal T'}_{z_j}$ that widens the definition of the set of isometries induced by $z_j$. Since ${\cal T'}_{z_j}$ contains in general $2^{n-k}$ isometries of $\O1n$ the corresponding superset ${\cal T}_{z_j}$ will contain $2^{n-k}$ equivalence classes, namely with (\ref{formula_C_t_general})
\begin{equation}
\label{formula_cal_T_j_def3}
{\cal T}_{z_j} = \cup_{\lambda \in {\cal T'}_{z_j}} C_\lambda \subset \OO{n}
\end{equation}
and with this definition we can prove:
\begin{MS_theorem}
\label{SAT_in_O(n)}
A given \SAT{} problem in $\myClg{}{}{\R^{n,n}}$ (\ref{formula_SAT_EFB_2}) is unsatisfiable ($S = 0$) if and only if the isometries induced by its clauses (\ref{formula_cal_T_j_def3}) form a cover for $\OO{n}$:
\begin{equation}
\label{formula_SAT_in_O(n)}
\cup_{j = 1}^m {\cal T}_{z_j} = \OO{n} \dotinformula
\end{equation}
\end{MS_theorem}
\begin{proof}
Let (\ref{formula_SAT_in_O(n)}) hold, by (\ref{formula_cal_T_j_def3}) and elementary properties of equivalence classes $\cup_{j = 1}^m {\cal T'}_{z_j} = \O1n$ and $S$ is unsatisfiable by Proposition~\ref{prop_SAT_in_On1}. Conversely let $S$ be unsatisfiable, the thesis follows from Proposition~\ref{prop_SAT_in_On1} and (\ref{formula_cal_T_j_def3}).
\end{proof}

This result gives an unsatisfiability test that verifies if the clauses $z_j$ induce a cover of $\OO{n}$. We just recall that $\OO{n}$ is a continuous group that form a compact, disconnected real manifold of dimension $n (n-1)/2$ and that the two connected components are respectively $\SO{n}$, with $\Identity$, and its coset given by $\OO{n}$ elements with determinant $-1$ \cite{Porteous_1981}.

An algorithm could test unsatisfiability checking if there are $\OO{n}$ elements not contained in $\cup_{j = 1}^m {\cal T}_{z_j}$ but now the continuity of $\OO{n}$ makes the situation more interesting with respect to the case of \O1n (\ref{formula_SAT_in_On1}).

We give an argument that supports this claim: exploiting a parametrization of $\OO{n}$ elements we can transform the sets ${\cal T}$ (\ref{formula_cal_T_j_def3}) induced by clauses into subsets of the parameter space (for example to subsets of $[0, 2 \pi)^{\frac{n(n - 1)}{2}}$ in the case of decomposition in Givens rotations, bivectors in Clifford algebra) and search after $\OO{n}$ elements not contained in $\cup_{j = 1}^m {\cal T}_{z_j}$ in this parameter space. Any outcome of this search would become, with (\ref{formula_C_t_general}) \cite{Budinich_2020}, an element of \O1n and a solution of the given \SAT{} problem. This hints a path to follow that, even if challenging, appears to be a non beaten track heading to unexplored territories and a worthy subject for future research.

\section{Conclusions}
\label{Conclusions}
We have shown how neatly Boolean algebra can fit in Clifford algebra and this allows to look at Boolean problems, and at \SAT{} in particular, from a different standpoint. A \SAT{} problem in CNF (\ref{formula_SAT_std}) is easily ported in Clifford algebra by (\ref{formula_SAT_EFB_2}) and its algebraic expansion is a sum of primitive idempotents corresponding to DNF, a sum of Boolean atoms. The algebraic expansion reproduces faithfully the corresponding Boolean one that exploits the distributive properties of logical AND and OR and, if on one hand validates the perfect fit of Boolean algebra in $\myClg{}{}{\R^{n,n}}$, on the other hand has no significant computational advantages to offer.

Nevertheless once a problem is firmly expressed in Clifford algebra we can exploit algebra properties and Theorem~\ref{theorem_SAT_unsat_thm} essentially says that a problem $S$ is unsatisfiable if and only if it has the maximally symmetric form of the scalars of the algebra. This yields a solution algorithm that is essentially a refinement of the well known Davis Putnam algorithm for \SAT{} \cite{Budinich_2017}. This result encourages to pursue further these studies at least since it provides a radically different interpretation of the Davis Putnam algorithm in term of projections in spinor spaces.

Successively we exploited the amazing one to one correspondence between seven apparently different topics, namely:
\begin{itemize}
\item the $2^n$ Boolean atoms of $n$ \Boov{}s,
\item the $2^n$ primitive idempotents of $\myClg{}{}{\R^{n,n}}$,
\item the $2^n$ simple spinors of the Fock basis $\myFockB$,
\item the $2^n$ $h$ signatures of simple spinors in EFB (\ref{formula_EFB_def}) \cite{Budinich_2016},
\item the $2^n$ totally null subspaces of maximal dimension of $\R^{n,n}$,
\item the $2^n$ elements of the abelian group \O1n,
\item the $2^n$ the equivalence classes (\ref{formula_C_t_general}) of the orthogonal group $\OO{n}$
\end{itemize}
to transform \SAT{} problem between any of these forms in the unifying setting of \myClg{}{}{\R^{n,n}}, with the hope that they can shed some light in the intriguing relations occurring between them.

Subsequent Theorem~\ref{SAT_in_O(n)} exploits pivotal Proposition~\ref{prop_2^n_MTNS}, a peculiar property of null subspaces of $\R^{n,n}$ thoroughly discussed in \cite{Budinich_2020}, showing that an unsatisfiable \SAT{} problem induces a cover of the continuous group $\OO{n}$ porting the arena to the real manifold of dimension $n (n-1)/2$ of the group.

We conclude recalling that any $t \in \OO{n}$ is represented by either an orthogonal matrix of $\R(n)$ or by a matrix of $\R(2^n)$ corresponding respectively to its vectorial and spinorial representations. We introduced a \SAT{} encoding that refers naturally to matrices of the spinorial representation but that could also be transferred to the, much more manageable, matrices of the vectorial representation this being another topic for future investigations.

\vspace{7cm}
\opt{x,std,AACA}{

\bibliographystyle{plain} 
\bibliography{mbh}
}
\opt{arXiv,JMP}{
%
%

%
%
}

\opt{final_notes}{
\newpage

\section{Good stuff removed from the paper}
\label{sec_Good stuff removed}
We underline that this procedure (of generating all \mysnqG{} elements of \mysetM{} at the beginning of Section~\ref{sec_O(n)_MTNS}) is remarkably similar to that of forming \SAT{} problems: we must choose, for each of $n$ \Boov{}s, whether we take it in plain or complemented form and this similarity is no chance since there is an one to one correspondance between the two procedures \cite{Budinich_2019}. The other way round as any spinor can be seen as the action of a succession of creation and destruction operators applied to a vacuum state, any logical expression can be built by the action of creation and destruction operators on any initial state.
\opt{margin_notes}{\mynote{mbh.note: to be developed}}%

} 

\end{document}